\documentclass[sigconf]{acmart}

\usepackage{graphicx}
\usepackage[show]{chato-notes}
\usepackage{mathtools}
\usepackage{booktabs}
\usepackage{bbm}
\usepackage{epsfig}
\usepackage{algorithm}
\usepackage{algpseudocode}
\usepackage{hyperref}
\usepackage{tikz}
\usepackage{microtype} %
\usepackage{xspace}
\usepackage{xcolor}
\usepackage{xfrac} %
\usepackage{url} %
\usepackage{hyphenat} %
\usepackage[hyperpageref]{backref}
\usepackage{cleveref}
\usepackage{siunitx}
\usepackage{enumitem}
\PassOptionsToPackage{hyphens}{url} %

\hypersetup{
    bookmarks=true,         %
    unicode=false,          %
    pdftoolbar=true,        %
    pdfmenubar=true,        %
    pdffitwindow=false,     %
    pdfstartview={FitH},    %
    pdftitle={My title},    %
    pdfauthor={Author},     %
    pdfsubject={Subject},   %
    pdfcreator={Creator},   %
    pdfproducer={Producer}, %
    pdfnewwindow=true,      %
    colorlinks=true,       %
    linkcolor=black,          %
    citecolor=black,        %
    filecolor=black,      %
    urlcolor=black,           %
    pagebackref=true
}

\newcommand{\R}[1]{\texttt{\small{r/#1}}}
\newcommand{\modelA}{socio-demographic features model\xspace}
\newcommand{\modelB}{socio-demographic model with topics\xspace}
\newcommand{\ModelA}{Socio-demographic features model\xspace}
\newcommand{\ModelB}{Socio-demographic model with topics\xspace}
\newcommand{\mA}{SD\xspace}
\newcommand{\mB}{SD+T\xspace}

\DeclareMathOperator{\logit}{logit}

\renewcommand*\backref[1]{\ifx#1\relax \else (Cited on #1) \fi}

 \title[Evidence of Demographic rather than Ideological Segregation in News Discussion on Reddit]{Evidence of Demographic rather than Ideological Segregation\\in News Discussion on Reddit}

\author{Corrado Monti}
\email{corrado.monti@centai.eu}
\orcid{0000-0001-6846-5718}
\affiliation{%
  \institution{CENTAI}
  \streetaddress{Corso Inghilterra 3}
  \city{Turin}
  \country{Italy}
  \postcode{10138}
}

\author{Jacopo D'Ignazi}
\email{jacopo.dignazi@isi.it}
\orcid{0000-0003-2843-5279}
\affiliation{%
  \institution{ISI Foundation}
  \streetaddress{Via Chisola 5}
  \city{Turin}
  \country{Italy}
  \postcode{10126}
}

\author{Michele Starnini}
\email{michele.starnini@centai.eu}
\orcid{0000-0002-9161-5339}
\affiliation{%
  \institution{CENTAI}
  \streetaddress{Corso Inghilterra 3}
  \city{Turin}
  \country{Italy}
  \postcode{10138}
}

\author{Gianmarco De Francisci Morales}
\email{gdfm@acm.org}
\orcid{0000-0002-2415-494X}
\affiliation{%
  \institution{CENTAI}
  \streetaddress{Corso Inghilterra 3}
  \city{Turin}
  \country{Italy}
  \postcode{10138}
}
\renewcommand{\shortauthors}{Monti et al.}

\copyrightyear{2023}
\acmYear{2023}
\setcopyright{acmlicensed}\acmConference[WWW '23]{Proceedings of the ACM Web Conference 2023}{May 1--5, 2023}{Austin, TX, USA}
\acmBooktitle{Proceedings of the ACM Web Conference 2023 (WWW '23), May 1--5, 2023, Austin, TX, USA}
\acmPrice{15.00}
\acmDOI{10.1145/3543507.3583468}
\acmISBN{978-1-4503-9416-1/23/04}

\ccsdesc[500]{Information systems~World Wide Web}
\ccsdesc[500]{Applied computing~Sociology}
\ccsdesc[300]{Human-centered computing~Social networking sites}

\keywords{Homophily, Socio-demographic factors, Polarization, News, Reddit}

\begin{document}

\begin{abstract}
We evaluate homophily and heterophily among ideological and demographic groups in a typical opinion formation context: online discussions of current news.
We analyze user interactions across five years in the \R{news} community on Reddit, one of the most visited websites in the United States.
Then, we estimate demographic and ideological attributes of these users.
Thanks to a comparison with a carefully-crafted network null model, we establish which pairs of attributes foster interactions and which ones inhibit them.

Individuals prefer to engage with the opposite ideological side, which contradicts the echo chamber narrative.
Instead, demographic groups are homophilic, as individuals tend to interact within their own group---even in an online setting where such attributes are not directly observable.
In particular, we observe age and income segregation consistently across years: users tend to avoid interactions when belonging to different groups.
These results persist after controlling for the degree of interest by each demographic group in different news topics.
Our findings align with the theory that affective polarization---the difficulty in socializing across political boundaries---is more connected with an increasingly divided society, rather than ideological echo chambers on social media.
We publicly release our anonymized data set and all the code to reproduce our results.\footnote{\url{https://github.com/corradomonti/demographic-homophily}}
\end{abstract}

\maketitle \sloppy

\graphicspath{{../fig/}{figures/}}

\section{Introduction}
\label{sec:intro}

Socializing across ideological boundaries has been reported to be increasingly difficult~\citep{iyengar2019origins}.
This phenomenon, known as \emph{affective polarization}, has been widely observed in the United States.
For instance, the fraction of Republicans and Democrats that would be unhappy if their child married a supporter of the opposing party has increased from 5\% in 1960, to %
more than 30\% in 2010~\citep{iyengar2012affect}.
This kind of polarization contributes to creating a harsh political climate, where citizens believe that other people's opinions are irrational and not based on facts~\cite{jackson2018fake}.
It has also been associated with the spread of harmful conspiracy theories, which can lead to dehumanization~\cite{harel2020normalization}, culminating in threats such as the emergence of far-right domestic terrorism~\cite{lowery2018united,michael2017rise}.

The causes of affective polarization are widely debated, and some studies have pointed at social media as a decisive factor~\cite{garrett2009echo}.
In this popular view~\cite{pariser2011filter}, social media---through recommendation algorithms, algorithmic filter bubbles, and self-sorting---would reinforce ideological separation, thus making it harder for an individual to engage with information from the opposite side.
While several studies have observed ideological homophily and echo chambers in some social media~\cite{conover2011political,garimella2018political,cota2019quantifying,cinelli2021echo}, %
in others interactions between opposing sides are the norm and not the exception~\cite{dubois2018echo,defrancisci2021noecho,treen2022discussion,cinelli2021echo}.
Some scholars have argued that online media has actually increased chances of engaging with the opposite ideological side~\cite{fletcher2018people}, and Internet usage explains only a very small share of polarization \cite{boxell2017greater}.

An alternative explanation for the increase in affective polarization lies in the role of demographic factors~\citep{mason2018one, bradley2022ethnic, phillips2022affective}.
Underlying material rifts and inequalities in society make gender, race, age, and wealth boundaries increasingly divisive, and also increasingly correlated to party affiliation and ideological standpoints.
In their seminal paper, \citet{lazarsfeld1954friendship} distinguished between status homophily, when similarity in social status fosters connections, and value homophily, when  similarity in beliefs and ideas tends to do so.
In this view, the existing status homophily within these groups would therefore be one of the main drivers behind value homophily and thus contribute to affective polarization~\cite{mcpherson2001birds}.

To assess these competing hypotheses, this study focuses on the interactions between different demographic groups in one of the main loci where opinions are formed and challenged: the discussion of news~\cite{kim1999news}.
In particular, the discussion of news on social media is of twofold interest.
First, it allows studying how specific demographic groups are connected in a context where they cannot directly observe each other's demographic conditions:
any measured effect is entirely due to differences in worldviews, interests, and thoughts expressed through writing. %
Second, it allows us to measure and compare, at the same time and in the same context, the likelihood of interaction between opposing ideological sides.
These reasons lead us to investigate Reddit, one of the most visited websites in the U.S., a public forum widely used to read and discuss news.
We focus on the \R{news} subreddit, the main Reddit community specifically dedicated to discussion of current news, with a focus on U.S. internal affairs.

Therefore, our main research question is \emph{whether social interactions in online news discussions on Reddit tend to be segregated by demographic boundaries, or by ideological echo chambers}.

To this aim, we first distinguish groups of users more likely to be part of a certain demographic group thanks to the method proposed by~\citet{waller2021quantifying}.
Then, we measure how much the amount of interactions observed between demographic groups differs from a carefully-crafted null model, where such attributes would not affect interactions.
We replicate this analysis separately for five years, from 2016 to 2020.

In all these experiments, we find similar effects.
First, that ideological echo chambers are absent: 
left-wing and right-wing users interact \emph{more} than expected by a null model of random interactions, while within-group interactions are less likely to happen.
This result suggests that ideological echo chambers do not necessarily appear in a typical context of opinion formation online. %
Second, interactions in news discussions on Reddit are unlikely to cross demographic boundaries, 
e.g., affluent users are more likely to interact among themselves than with less-affluent users.
In other words, we observe \emph{segregation} along demographic boundaries, in the sense of ``restriction of contacts between various groups''~\citep{freeman1978segregation}, in line with previous literature on social networks~\cite{hofstra2017sources}.
Finally, we demonstrate that demographic homophily is not simply driven by common interests, such as discussions on the same topics.
While indeed different demographic groups have different interests in the various news topics, their segregation goes beyond this simple explanation.
Most importantly, this segregation happens without users being able to directly observe the demographic attributes, and without the mediating effects of, e.g., geographic segregation.

Our results, therefore, support the idea that not only echo chambers might not be an intrinsic characteristic of social media~\citep{cinelli2021echo},
but that the separation of demographic groups is observable during opinion formation online:
fundamental societal divisions might be reflected also in social media. %

\section{Background and related work}
\label{sec:related}

The polarization of opinions, with respect to politics and controversial topics in general~\citep{pew2017partisan}, has been reported to be on the rise in recent years~\citep{jacobson2016polarization}.
Online social media, where billions of people express their opinions daily, offer a scalable and inexpensive way to quantify this phenomenon.
The literature on this topic is too vast to review here, so we briefly touch on a few landmark studies.
In a pioneer study, \citet{conover2011political} showed that on Twitter, a network of retweets about the 2010 US midterm elections exhibits a highly segregated partisan structure, with extremely limited connectivity between left- and right-leaning users.
Since then, their result has been replicated several times, on different topics, and in various contexts~\citep{garimella2016quantifying,cossard2020falling,garimella2017long,morales2015measuring,darwish2019quantifying}.
Opinion polarization has also been detected on Facebook and YouTube, which may foster selective exposure of users toward content they like, thus limiting the exposure to diverse content~\cite{bessi2016users,an2014partisan,bakshy2015exposure,harel2020normalization,quattrociocchi2016echo,schmidt2018polarization}.
Echo chambers, situations where users have their beliefs reinforced due to repeated interactions with like-minded peers, have been proposed as one of the determinants for polarization on social media, and as a threat to democracy~\cite{sunstein2009republic}.

The investigation around the presence of echo chambers on social media has recently spurred a rich literature and a vivid debate among researchers.  
On Twitter, the content consumption and production of users reveal a similar political leaning, so that users are generally exposed to political opinions that agree with their own~\cite{garimella2018political}.
On the same platform, the presence of echo chambers in the debate about the impeachment of former Brazilian President Dilma Rousseff has been shown to alter the information diffusion between the two sides, supporters and opponents of the impeachment~\citep{cota2019quantifying}.
On Facebook, \citet{quattrociocchi2016echo} showed that users form polarized groups that share two distinct narratives,  related to conspiracy theories and science, respectively.
On the same platform, the content consumption about vaccines is skewed by echo chambers, while opinion polarization increased over the years~\citep{schmidt2018polarization}.
Finally, \citet{cinelli2021echo} compared the presence of echo chambers around several controversial topics across four social media platforms (Twitter, Facebook, Reddit, and Gab), highlighting that Facebook shows higher segregation of news consumption than Reddit.
Interestingly, several works challenge the harmfulness of the phenomenon~\citep{sunstein2005societies}, or the very existence of echo chambers in different contexts~\citep{dubois2018echo,flaxman2016filter,guess2018avoiding}.
For instance, on Reddit, Trump and Clinton supporters show a preference for cross-cutting political interactions between the two communities rather than within-group interactions, thus contradicting the echo chamber narrative~\cite{defrancisci2021noecho}.
The current work finds a similar result in a different context (news discussion), and at the same time finds evidence for demographic segregation on social media.

Since online discussion can be a driving factor of opinion formation~\citep{kim1999news}, the search for the causes of polarization has ultimately led to questions being asked on how people interact on the base of their socio-demographic extraction, culture, and ideology.
Demographic segregation in physical spaces has been the subject of study for several decades~\citep{schelling1969models}, and has been shown to have increased in modern times~\citep{logan2017national}, mostly due to racial sorting.
The idea that this sorting is the main cause for increased political polarization has been advanced~\citep{bishop2009big,levendusky2009partisan}, although it has also been criticized~\citep{abrams2012big}.
In this context, affective polarization has been introduced as a measure of the attitude of individuals with opposite ideologies toward each other~\citep{iyengar2012affect}.
\citet{iyengar2015fear} have shown how partisan-wise polarization also leads to increasing animosity (and even loathing) across party lines.
Indeed, the social identities of people have grown increasingly aligned with their partisan identity~\citep{iyengar2019origins,mason2018one}. 
Nonetheless, recent research has found how affective polarization can be influenced by factors such as ethnicity~\citep{bradley2022ethnic} and age~\citep{phillips2022affective}, thus suggesting that socio-demographic and cultural differences might be an important driving force in social interactions.
This idea is corroborated by observations of how social interactions are strongly homophilic with respect to socio-demographic extraction~\citep{mcpherson2001birds, lazarsfeld1954friendship}.
All these findings suggest that more attention is needed to the role of socio-demographic and cultural features, as pivotal factors in influencing individual social choices, and consequently opinion formation and polarization.

To address this gap, we employ graph models that allow studying the interplay between node features and link formation.
Our choice of model is driven by a few constraints.
For instance, the well-known stochastic block model~\cite{nowicki2001estimation}, in its typical form, only allows a node to belong to exactly one class, while we are interested in modeling the presence of more features on nodes at the same time.
Also, we need a model able to capture feature-feature relationships that are not limited to homophily.
Thus, we opt for the feature-feature model initially proposed by \citet{miller2009nonparametric} as a way to learn features from data, and then further studied as a generative network model~\citep{boldi2016network}, and as a tool to infer feature-feature relations~\citep{monti2017estimating,boldi2016llamafur}.
In the present work, we use this network model, embedded inside a logistic regression, to evaluate the statistical significance of the inferred correlations.
Then, we extend it to control for varying degrees of interest in news topics by different demographic groups.

\section{Data}
\label{sec:data}

This section describes the data sources we use as input for our analyses.
First, Reddit and the \R{news} interaction graphs we built.
Then, the socio-demographic features extracted from Reddit participation data.
Finally, the news topics, that we employ as confounders.

\subsection{Reddit interaction graph}
Reddit is a social content aggregation website that has consistently ranked among the top ten most visited websites in the United States over the past years.\footnote{\url{https://en.wikipedia.org/wiki/Reddit}}
It is organized in topical communities, called \textit{subreddits}, centered around a variety of topics.
In particular, \R{news}, the focus of the current work, is dedicated to discussing ``news articles about current events in the United States and the rest of the world''.\footnote{\url{https://www.reddit.com/r/news}}
Users can pseudonymously post submissions in these subreddits, and comment on other submissions and comments, thus creating a tree structure for the overall discussion.

Differently from other social media such as Facebook or Twitter, Reddit's homepage is organized around subreddits rather than user-to-user relationships.
As such, the subreddits chosen by a user represent the main source of the information they consume on the website.
We collect public data from Reddit by using Pushift~\citep{baumgartner2020pushshift}.

\label{sec:data_user}

Since Reddit data is massive---around $10^7$ comments each year---we reduce our datasets to what we consider the most relevant interactions. 
First, we choose users active on the subreddit \R{news} from 2016 to 2020, included. 
Then, we select users by applying the following criteria: a minimum activity on \R{news}, a minimum activity on the whole Reddit, and that they are not likely to be a bot.
In particular, a selected user must have written ($i$) at least 25 between submissions and comments on \R{news} in a given year,\footnote{We also perform all the experiments with a threshold of 10 messages per year: this choice does not significantly alter the results.} and ($ii$) submissions to at least 5 different subreddits in a given year.
These empirically-determined cutoffs serve a twofold purpose: to select the most active users to avoid sparsity issues and to reduce computational costs. 
Moreover, as we show next, participation in other subreddits is necessary for estimating socio-demographic features.
To exclude news bots, we filter out users who either ($i$) made submissions on \R{news} but no comments, ($ii$) belong to a list of known Reddit bots (e.g., automoderator), ($iii$) have posted in more than 50 subreddits each month, or ($iv$) have the string 'bot' in their username.
    
Starting from the selected set of users $V$, we create an interaction graph $G$ by considering the users as nodes and defining an arc $u \rightarrow v$ if a user $u$ replies to a message posted by user $v$.
\Cref{tab:dataset} reports the number of selected users and social interactions.

\begin{table}[t]\label{datatab}
    \caption{Number of nodes and edges for the considered interaction graph of \R{news} in each year.}
    \label{tab:dataset}
    \vspace{-\baselineskip}
    \begin{tabular}{lrrrrr}
    \toprule
    Year        & \textbf{2016}    & \textbf{2017}    & \textbf{2018}    & \textbf{2019}   & \textbf{2020} \\ 
    \midrule
    N. nodes & \num{27976}   & \num{34060}   & \num{31997}   & \num{21225}  & \num{29045} \\
    N. edges & \num{1166076} & \num{1390243} & \num{1221779} & \num{793569} & \num{1067614} \\
    \bottomrule
    \end{tabular}
    \vspace{-\baselineskip}
\end{table}

\subsection{Socio-demographic features}%
\label{sec:feats:embedding}

To extract the socio-demographic features of the users, we use existing subreddit features obtained by embedding social interactions on Reddit.
\citet{waller2021quantifying} assign a score for different demographic axes (such as age, gender, and affluence) and social axes (political leaning) to each subreddit.
These scores are based on a set of well-thought pairs of seeds for each axis (e.g., \R{teenagers} and \R{redditforgrownups} for age), which are then used to construct the scores via embeddings, and finally validated with external sources of information.
We project these scores onto the users, and define the score $F_u$ of user $u$ with respect to feature $F$ as the weighted average of all their submissions in each subreddit,
\begin{equation}
    \label{eq:score}
	F_u=\frac{\sum_s N_{u,s} F_s}{\sum_s{N_{u,s}}},
\end{equation}
where $s$ is a subreddit, $F_s$ is the feature score of subreddit $s$, and $N_{u,s}$ is the number of submissions by user $u$ in subreddit $s$.
For instance, if a subreddit is leaning towards `male' in the gender axis, (i.e., it is more likely to be frequented by males), this scoring method will assign a larger probability of being male to a user who writes often on that subreddit.

\Cref{eq:score} allows us to characterize each user by a set of continuous scores that represent their position within each feature axis.
To use these features in a logistic regression model, we quantile-normalize these values among the selected users, and---to streamline our analysis---we binarize them, labeling only the ones with the highest and lowest quartile for each axis.
For instance, we label a user as male-leaning on the gender axis if their score falls in the lowest quartile of the gender axis (similarly, as female-leaning if in the highest quartile).
We stress that a score indicating a female-leaning user does not imply that such a user is necessarily female, as Reddit is known to be participated by more men than women.
Rather, it indicates a user that frequents subreddits is more likely to be participated by women.
Conversely, a higher score on the age axis simply indicates that the user is more likely to be older relative to the Reddit user base, which is quite young~\cite{barthel2016reddit}. %
Similarly, being in the lowest or highest quartile on the partisan axis does not imply ideological extremeness, but rather, statistical confidence about their partisan affiliation given their Reddit activity.
Users with very sparse or contradictory activity will instead be placed around the median: as such, we will consider them as a baseline.

\begin{figure}[btp]
    \centering
    \includegraphics[width=0.8\columnwidth]{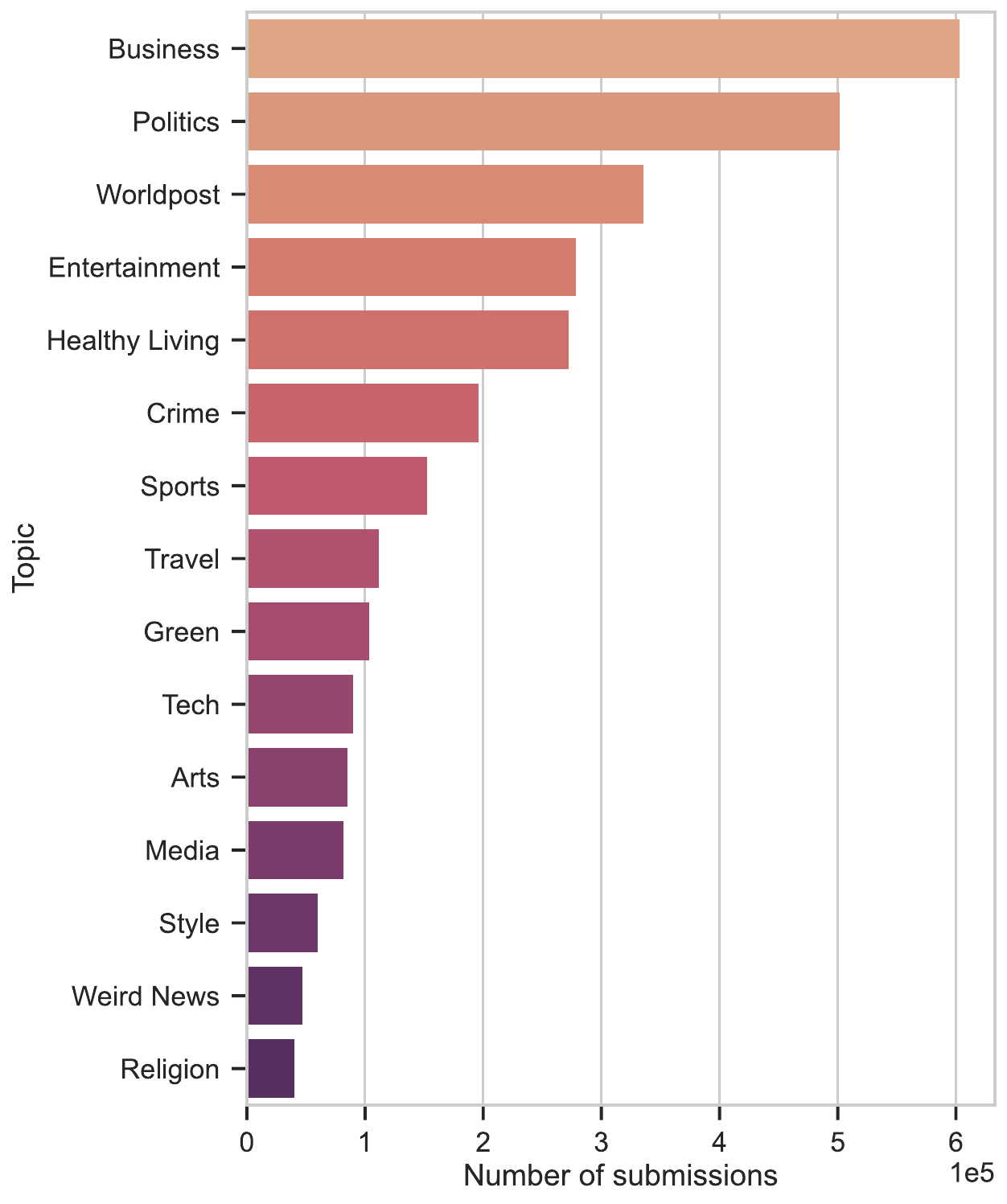}
    \vspace{-\baselineskip}
    \caption{Distribution of topics across the selected submissions of \R{news}.}
    \label{fig:topics}
    \vspace{-\baselineskip}
\end{figure}

\subsection{Topic classifier}
\label{sec:topic-classifier}

To check whether the effects of socio-demographic attributes on interactions are simply a byproduct of a varying degree of interest in news topics, we need to classify posts in \R{news} according to their news topic.
To do so, first, we train a state-of-the-art supervised topic detection classifier on news article headlines.
Then, we apply this classifier to the news appearing in \R{news}, so to obtain a mapping of posts into topics.

To train the supervised classifier we replicate results from the best-performing news topic detection model from Kaggle,\footnote{\url{https://www.kaggle.com/code/hengzheng/news-category-classifier-val-acc-0-65/}} which we describe below.
As a training set, we use a publicly available data set gathered by Misra and Grover~\cite{misra2022news,misra2021sculpting}.
This data set collects news headlines from the Huffington Post, each associated to the topic assigned by the editorial staff according to their topic categorization.
Since the posts in \R{news} are links to newspaper articles, this data source is homogeneous with the one we collect. 
The data set contains \num{210294} news headlines published between 2012 and 2022 and 40 topics.
Then, we use GloVe embeddings~\cite{pennington2014glove} with a dimension of 100 to encode the words identified after tokenization.
Finally, we employ a long short-term memory (LSTM) attention neural network to perform classification by using these embeddings.
In $10$-fold cross-validation on the news data set, this classifier obtains an average accuracy of $65\%$, which is considerably high considering that this is a 40-class classification problem.

After this training phase, we apply the resulting classifier to the \num{3304044} news headlines contained in our \R{news} data set and obtain a mapping of posts into topics.
Since each social interaction takes place under a post, each link in our interaction network is associated to a single topic.

Figure~\ref{fig:topics} shows the 15 most popular topics in our data set.
These 15 topics account for $90\%$ of our data set; therefore, we exclude the remaining 27 topics from our analysis (the largest excluded topic accounts for $1.1\%$ of the posts).
The most popular topic is \emph{Business} ($18.2\%$ of posts), followed by \emph{Politics} ($15.2\%$) and \emph{Worldpost} ($10.2\%$), a topic category related to news events taking place outside the United States.
The topic \emph{Entertainment} ($8.4\%$) pertains mostly to news about movies and music.
\emph{Healthy Living} ($8.2\%$) collects news regarding healthy lifestyles and habits.
Other popular topics---beside self-explanatory ones such as \emph{Crime}, \emph{Tech}, and \emph{Religion}---include \emph{Green}, related to environment and climate change;
\emph{Media}, about TV and movies; \emph{Style}, about fashion and beauty care; and \emph{Weird News}, collecting strange-but-true headlines, from UFOs to pets.
To further exemplify the topics, we report some headlines from our Reddit dataset in \Cref{tab:news-examples}.

\begin{table}[tb]
    \caption{Sample of news headlines found in our Reddit data set for each considered topic.}
    \label{tab:news-examples}
    \vspace{-.5\baselineskip}
    \centering
    \footnotesize    
\begin{tabular}{lp{0.8\columnwidth}}
\toprule
         Topic &                                                                                         News headline \\
\midrule
      Business &                                Goldman's Investment Bank to Increase its Hiring of Black Employees \\
      Politics &                                Trump attacks Schiff: `He lies awake at night shifting and turning' \\
     Worldpost &                                     From China to Italy, Coronavirus leaves a trail of devastation \\
 Entertainment &                                                                       Oscar 2020 Winners Full List \\
Healthy Living &                                   Nursing Homes, Racked by the Virus, Face a New Crisis: Isolation \\
         Crime & Video Shows Missouri Cop Run Over Fleeing Suspect With Car, Then Tackle Him As He Screams \\
        Sports &                                                   Moloney Vs Baez - Boxing, June 25, 2020, On ESPN \\
        Travel &                         Double rainbows form in skies over Delhi, surrounding areas  \\
         Green &  YouTube Has Been `Actively Promoting' Videos Spreading Climate Denialism, According to New Report \\
          Tech &                                    Parallels and Google team up to bring Windows apps to Chrome OS \\
          Arts &                                                   In Montana, the Art of Crafting Fly-Fishing Rods \\
         Media &                                               Jim Lehrer, Longtime PBS NewsHour Anchor, Dies at 85 \\
         Style &                                                                                     New year gifts \\
    Weird News &                                       Cops: McDonald's mask scofflaw breaks window, steals panties \\
      Religion &                                              Join Wajahat Ali to discuss Ramadan during a pandemic \\
\bottomrule
\end{tabular}
\normalsize
\vspace{-\baselineskip}
\end{table}

\section{Model}
\label{sec:model}

This section introduces the two logistic regression models we develop to estimate the relationships among the socio-demographic groups in Reddit \R{news}.
Each one is built upon a network null model where the dependent variable indicates whether a pair of nodes have interacted on the subreddit.
The first one, dubbed \modelA (\mA), uses only independent variables derived from the socio-demographic features.
The second model, the \modelB (\mB), investigates whether such relationships hold after controlling for different interest by each demographic group in each news topic.

\subsection{\ModelA}
\label{sec:modelA}

The goal of the \mA model is to estimate which pairs of socio-demographic features foster or inhibit interactions among Reddit users.
To do so, we rely on the following logistic regression approach, inspired by the feature-feature graph model~\citep{miller2009nonparametric, monti2017estimating,defrancisci2021noecho}.
Consider a set of features $F$ and a directed graph $G=(V, E)$.
Each node $v \in V$ is characterized by a binary vector of features $x_v \in \{ 0, 1 \}^{|F|}$.
Then, for a node pair $(u, v) \in V \times V$, let the dependent variable $y$ be equal to $1$ if the two nodes interact (i.e., $(u, v) \in E$) and $0$ otherwise.
Our model estimates the relationship of the dependent variable $y$ with the vectors of independent variables (features) of the two nodes, $x_u$ and $x_v$.
More specifically, we consider all the possible pairs of features $F \times F$: the hypothesis is that the likelihood of observing an interaction from $u$ to $v$ can be affected by node $u$ having feature $h$ and node $v$ having feature $k$, for any feature pair $(h, k) \in F \times F$.
This model can be expressed by a feature-feature matrix $W \in \mathbb{R}^{|F| \times |F|}$, where each element $W_{h,k}$ indicates the log odds ratio for nodes with feature $h$ to interact with nodes with feature $k$.
A positive weight indicates that nodes with features $h$ are more likely to interact with nodes with features $k$ with respect to the probability of random interactions given by the null model (see below).
Conversely, a negative element $W_{h,k}$ indicates that this specific pair of features inhibits interactions.

To convert this framework into a logistic regression model, we apply an outer-product kernel~\citep{monti2017estimating} to $x_u, x_v$.
In other words, the independent variables of the logistic regression are all the elements of $x_u \otimes x_v$; that is, each pair of nodes $u, v$ is associated to $|F| \cdot |F|$ variables such that the element $h, k$ is $1$ if node $u$ has feature $h$ and node $v$ has feature $k$.
Therefore, the model can be expressed as
\begin{equation}
\label{eq:modela}
    \logit(y_{u, v}) =
    \beta_0 + x_u^\intercal W x_v =
    \beta_0 + \sum_{h,k \in F} x_{u, h} W_{h, k} x_{v, k}.
\end{equation}
where $\beta_0$ indicates the intercept, which adjusts the model according to the average probability of an interaction between two nodes.

In our context, the directed graph $G=(V, E)$ is the graph of interactions in Reddit \R{news} in a given year, as defined in \Cref{sec:data_user}. 
The features used in this model are the socio-demographic variables defined in \Cref{sec:feats:embedding}.
The set $F$ is therefore composed by features: Young, Old, Male, Female, Poor, Rich, Left-leaning, and Right-leaning.
We assign such features to nodes as explained in \Cref{sec:feats:embedding}, so that $x_{u, k}$ is $1$ iff node $u$ is in the quartile of considered users more likely to be in socio-demographic group~$k$.
This way, the logistic regression model fits a given sequence of pairs $(y_{u,v}, x_u \otimes x_v)$, and obtains an estimate of the feature-feature matrix $W$.
Such a matrix indicates which socio-demographic groups are more likely to interact.

The construction of the sequence of node pairs is crucial, as it represents the null model. %
We would like this null model to be the random graph obtained by reshuffling the links of the original network, while preserving the in- and out-degree of each node---that is, a configuration model~\citep{molloy1995critical}:
this way, the activity of each user is properly taken into account.
Therefore, we define a balanced sequence of pairs, i.e., we sample a number of negative examples equal to the positive ones.
We do so by, first, including all positive examples (i.e., the links $E$); then, as negative examples, we choose node $u$ with probability proportional to their activity (i.e., the number of posted comments), and node $v$ with probability proportional to their attractiveness (the number of received comments)~\cite{defrancisci2021noecho}.
If the obtained pair $(u, v)$ is a link, we discard it. This way, the null model expresses the probability of considering a pair of nodes as the product of two independent probabilities---the probability that a node initiates an interaction, and the probability that a node receives it.
The estimated matrix $W$ will therefore represent how the socio-demographic features make the observed links deviate from this null model.
Since the estimation of matrix $W$ is obtained from a logistic regression model, it is also accompanied by a statistical test able to tell if each element $W_{h,k}$ is significantly different from zero or if it might be spurious.

\subsection{\ModelB}
\label{sec:modelB}

One might wonder whether the status homophily we see is just the result of shared interests.
That is, similar people are interested in and comment on similar topics, and therefore are more likely to interact simply because they are drawn towards the same posts.
In other words, we wish to assess whether the topic of the post (news) acts as a mediator~\cite{pearl2009causality}, or conversely, whether there are any residual effects of the socio-demographic features once we control for the topic, which may act as a confounder~\cite{pearl2009causality}.

To this aim, we augment the \mA model to account for the topic of the post by introducing a feature-topic matrix $Q \in \mathbb{R}^{|F| \times |T|}$, where $T$ is the set of possible topics defined in \Cref{sec:topic-classifier}.
Each entry $Q_{h,t}$ of the matrix represents the log odds ratio that a user with feature $h$ is interested in topic $t$ (and therefore comments on it).
To formalize the model we make use of a binary indicator (one-hot) vector $\tau^p \in \{0,1\}^{|T|}$ which for each post $p$ (on which the interaction between $u$ and $v$ happens) has value $1$ only for the topic of the post, and is $0$ otherwise.

We integrate this feature-topic component $Q$ into the model expressed by \Cref{eq:modela} by assuming that, beside the feature-feature effects expressed by $W$, the probability of observing a link $u \rightarrow v$ on topic $t$ is also affected by the interest of user $u$ in topic $t$, and the interest of user $v$ in the same topic.
Thus, we can formally express this model is as follows:
\begin{equation}
\label{eq:modelb}
    \begin{split}
    \logit(y_{u, v}) = \;
    & \beta_0 + x_u^\intercal W x_v + x_u^\intercal Q \tau^p+ x_v^\intercal Q \tau^p = \\
    & \beta_0 + \sum_{h,k \in F} x_{u, h} W_{h, k} x_{v, k} +
    \sum_ {h \in F} x_{u,h} Q_{h,t} + \sum_{k \in F} x_{v,k} Q_{k,t},
    \end{split}
\end{equation}
where $t$ represents the index of the topic of post $p$ in $T$. 

The matrix $Q$ is the same for both the initiator and the receiver of the interaction, i.e., it is shared between $x_u$ and $x_v$.
In fact, it is also possible to write the topic terms of Equation~\eqref{eq:modelb} as $(x_u+x_v)^\intercal Q \tau^p$, where $(x_u+x_v)_k$ indicates therefore the number of participants in the interaction with feature $k$.
As such, $Q$ represents a non-directional effect that captures the interest of socio-demographic groups in each news topic.
In addition, the increase in complexity of the model is limited: we need to estimate only an additional $|F| \cdot |T|$ parameters.
If the status homophily we measure is due simply to similar interests, we would expect the parameters $Q$ to capture most of the effect, and therefore make $W$ not significant.

\section{Results}
\label{sec:results}

We now use the data we gathered in combination with these models to answer our research question.
We articulate our main research goal into the following research questions:

\begin{enumerate}[label={RQ\arabic*}]
\item Are there ideological echo chambers in \R{news}? I.e., are interactions among users on the same political side more likely?
\item Is there evidence of demographic segregation? I.e., are interactions among users within the same demographic group more likely?
\item Do our findings on RQ1 and RQ2 hold after controlling for varying degrees of interests in news topics by different demographic groups?
\item Which news topic is each demographic group more interested in?
\end{enumerate}

First, we use the \mA model to assess whether interactions are more homophilic or heterophilic with respect to ideological (RQ1) and demographic attributes (RQ2). %
Then, we use the \mB model to investigate whether such results hold after controlling for the varying degree of interest in news topics (RQ3). %
Finally, we analyze the results of the latter model to study which news topics are more attractive for the different socio-demographic groups (RQ4). %

\subsection{\mA model}
\label{sec:modelA-results}

\begin{figure}[tbp]
    \centering
    \includegraphics[width=0.9\columnwidth]{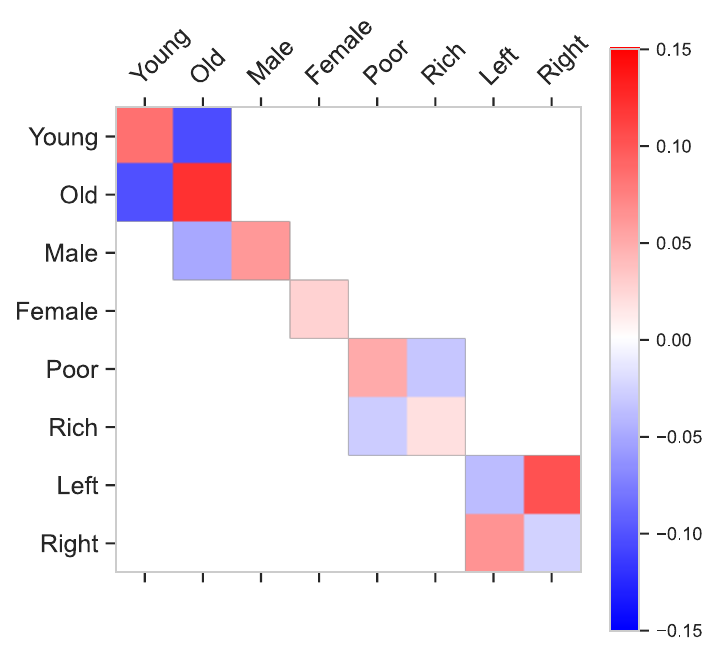}
    \vspace{-\baselineskip}
    \caption{Logistic regression coefficients for interactions between all pairs of demographic features obtained for the \R{news} subreddit in 2016 by the \modelA. We show only coefficients that are significant in at least 4 years out of 5.}
    \label{fig:full-matrix}
    \vspace{-\baselineskip}
\end{figure}

We apply the \mA model to the interaction graph, using the $8$ socio-demographic features we defined, separately for each considered year (2016 to 2020).
We obtain an estimate of the feature-feature matrix $W$ for each year.
As a first representation of our results, \Cref{fig:full-matrix} shows the matrix of one year (2016), but shows only the coefficients that are statistically significant in at least four years out of five.
More specifically, in this context, we define a coefficient $W_{h, k}$ as significant if the null hypothesis of $W_{h,k}=0$ is rejected ($p$-value $< 0.05$), and the coefficient $W_{h,k}$ has the same sign across all the considered years.
The base case for the model is a zero-valued feature vector, i.e., a user who is not in the top or bottom quartile on any socio-demographic axis.

Observing \Cref{fig:full-matrix}, the most striking characteristic is the diagonal---i.e., each socio-demographic group tends to interact significantly more, or less, with members of the same group.
In other words, each pair of features display a consistent effect in terms of heterophily and homophily.
Remarkably, this effect is heterophily for the political leaning and homophily for the demographic features.
Left-leaning and right-leaning users interact significantly more across the boundary, and significantly less within their respective group, w.r.t. the null model.
Interactions with the opposite ideological camp are the norm and not the exception in \R{news}:
our answer to RQ1 is therefore negative.
Conversely, regarding age, gender, and affluence, users tend to interact significantly more \emph{within} their demographic group.
For age and affluence, in particular, we observe a consistent tendency of users to segregate: belonging to different groups inhibits interactions between users, thus leading to a positive answer to RQ2.

\begin{figure}[tbp]
    \centering
    \includegraphics[width=\columnwidth]{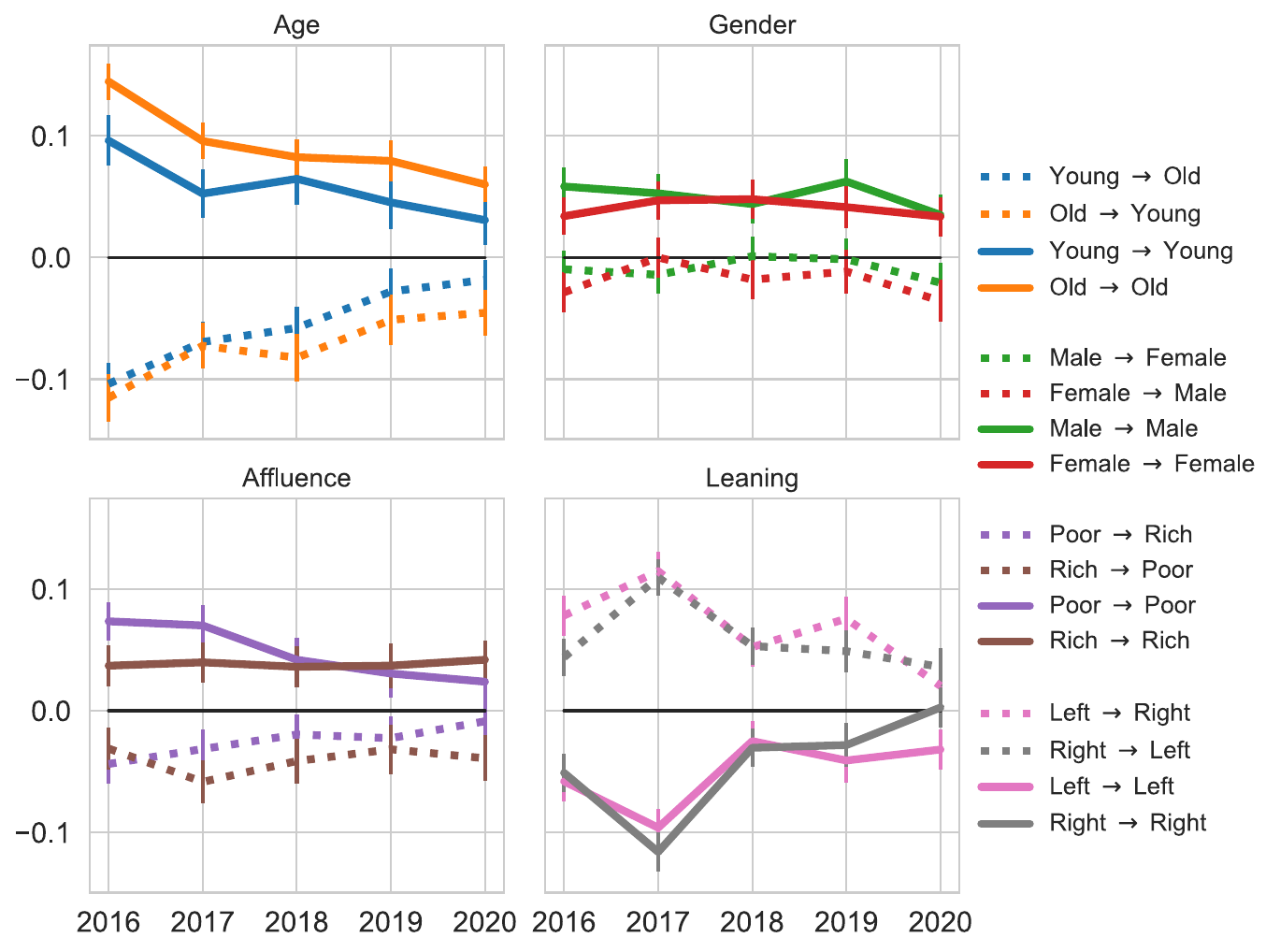}
    \vspace{-\baselineskip}
    \caption{Logistic regression coefficients for interactions between socio-demographic features obtained (independently) for each year in the \R{news} subreddit in the \mA model.}
    \label{fig:W-modelA}
    \vspace{-\baselineskip}
\end{figure}

Since we observe that most of the significant coefficients belong to interactions between pairs of demographic features, we further investigate them in \Cref{fig:W-modelA}. %
In particular, we show the block-diagonal of the weight matrix for each socio-demographic axis, one per subplot.
On the x-axis we report the year, on the y-axis the value of the $W$ coefficient (error bars represent the $95\%$ CI), solid lines represent within-class interactions, and dotted lines represent cross-class ones.
Again, the most important global pattern that emerges from the figure is that, for demographic features, within-class interactions are more frequent than expected, while the opposite is true for political leaning.
Overall, interactions on \R{news} are characterized by \emph{status homophily} and \emph{value heterophily}.

The pattern is consistent and statistically significant over the years, with the exception of the gender axis, where the within-class interactions are not far from the null model.
We also see a general trend towards convergence as we get closer to 2020 (especially pronounced in the age and leaning axes).
This effect might be due to the highly polarizing 2016 US presidential election, which saw an important participation on Reddit, while in 2020 Trump supporters were banned from Reddit, and resorted to alternative platforms such as Gab and Parler~\citep{zhou2018gab,aliapoulios2021early}.

\subsection{\mB model}
\label{sec:modelB-results}

\begin{figure}[tbp]
    \centering
    \includegraphics[width=\columnwidth]{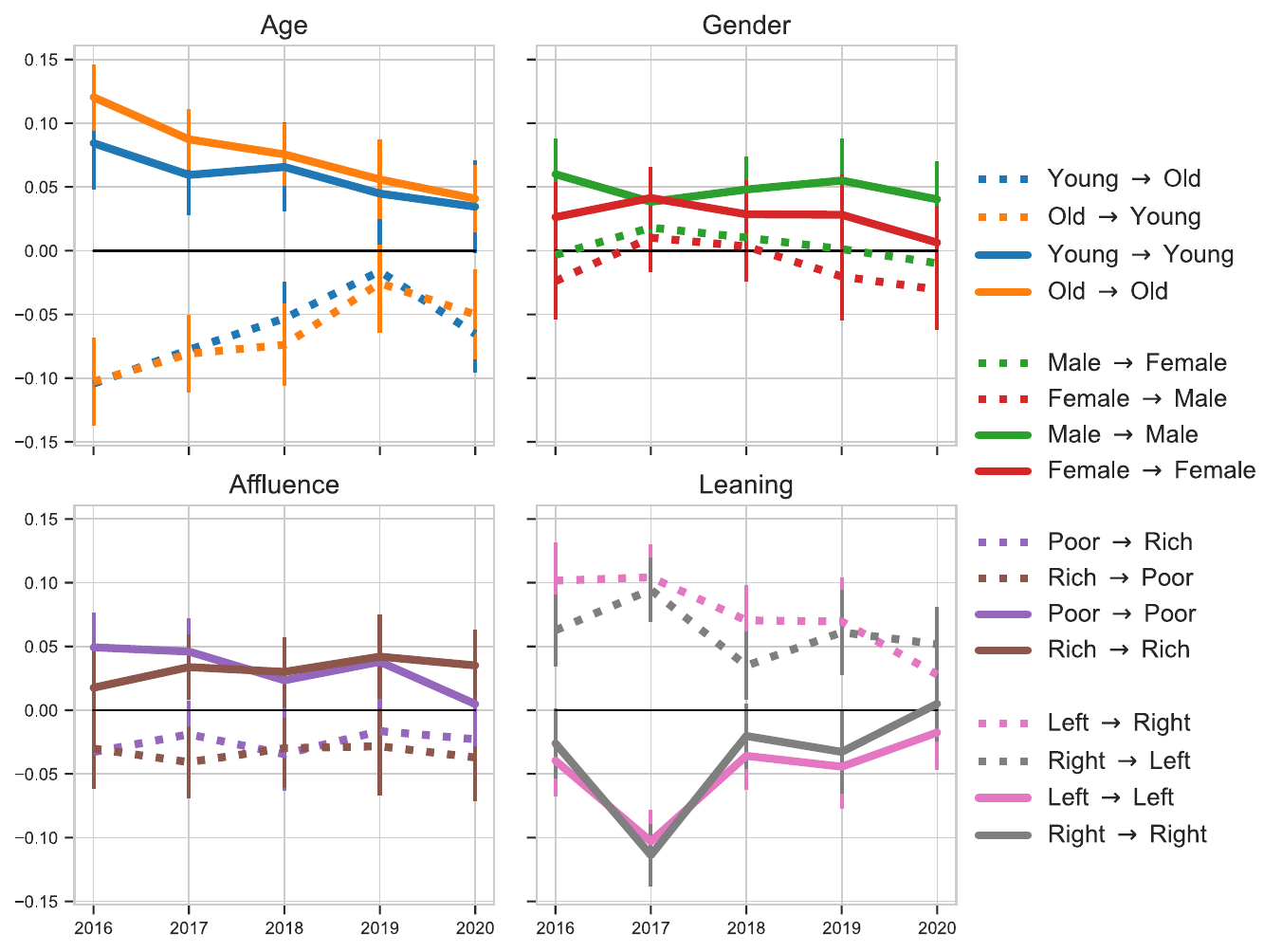}
    \vspace{-\baselineskip}
    \caption{Logistic regression coefficients for interactions between demographic features obtained (independently) for each year in the \R{news} subreddit in the \mB model.}
    \label{fig:W-modelB}
    \vspace{-\baselineskip}
\end{figure}

We now move to the \modelB, which controls for the topic of the discussion where the interaction takes place.
\Cref{fig:W-modelB} shows the weights of the matrix W for the logistic regression model of \modelB (again, the block-diagonal for each axis).
The main result is that the patterns do not change much from what we observed in \Cref{fig:W-modelA}, which leads to a positive answer to RQ3.
Indeed, we observe again status homophily and value heterophily, as within-class interactions along the age, gender, and affluence axes are more likely than expected, while cross-class ones are more likely for political leaning.
The same trends are present also in this case: convergence towards zero as we approach 2020, and non-significance of the cross-class gender terms.
Overall, controlling for the topic does not alter the results of the model, although it considerably improves its fit.
This fact indicates that the topic acts as a confounder and not as a mediator:
while the homophily and heterophily effects are not caused solely by the topic of the post, interactions are nonetheless partially driven by the different interests of demographic groups in different topics. %

\label{sec:modelB-topics}

\begin{figure*}[tbp]
    \centering
    \includegraphics[width=\textwidth]{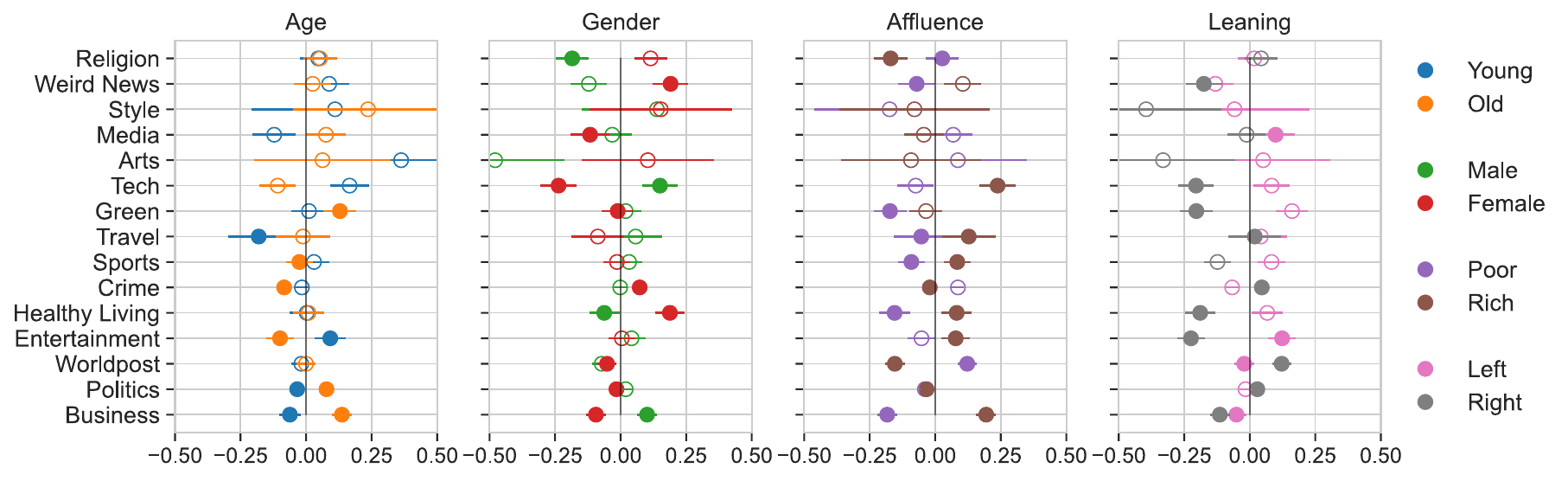}
    \vspace{-\baselineskip}
    \caption{Logistic regression coefficients (and $95\%$ CI) for interactions between demographic features and topics obtained for the \R{news} subreddit in 2016 by the \mB model. %
    Filled disks indicate associations significant in at least 4 years out of 5.}
    \label{fig:Q-modelB}
    \vspace{-\baselineskip}
\end{figure*}

Therefore, we look at the coefficients of the topic-class interactions encoded in matrix $Q$ of \modelB in order to answer RQ4.
\Cref{fig:Q-modelB} shows the weights of the matrix for one year (2016, the other years are qualitatively similar).
Each row represents a topic, each column an axis, and the value for each coefficient is full if it is statistically significant at least $4$ out of $5$ years, otherwise, it is hollow.
The error bars represent the $95\%$ CI of the estimates. A few interesting observations can be made. %

On the age axis, younger people are more interested in arts, technology, and entertainment, while older users are more interested in business and politics.
Older users are also more interested in green/environment topics, which might seem counter-intuitive, but can be understood by looking at the demographic of Reddit's user base, where older users are in their 30's or 40's, and younger ones are in their teens.
On the gender axis, typically male users show more interest in technology and business, but also lifestyle.
More female users are interested in religion, weird news, healthy living, and lifestyle.
On the affluence axis, low-income users are interested in religion and world news, while in general, they seem less interested in environment, travel, healthy living, business, and sports.
Conversely, richer users are interested in technology, travel, business, and to a lesser degree in healthy living and entertainment.

Finally, on the political leaning axis, we see a strong effect for right-wing users who are clearly less interested in arts and lifestyle news, and also show disinterest in several other topics including technology, environment, healthy living, entertainment, and business.
The only sizeable positive coefficients are for world news and crime.
Left-leaning users, instead, show an interest in arts, media, and entertainment.

\section{Discussion}
\label{sec:discussion}

We have studied social interactions among Reddit users in an opinion formation context: the discussion of news on the subreddit \R{news}.
By looking at the process through a socio-demographic lens, we have uncovered informative and consistent patterns.
Specifically, the interaction network displays clear status homophily, whereby users similar in age, gender, and affluence, are more likely to interact.
In addition, we find value heterophily, as users with opposing political leaning are also more likely to interact.
Importantly, these patterns are robust to controlling for the interest in the topic of the news:
the preference of a socio-demographic group for a news topic is not sufficient to explain the bias we observe.
Even inside the same news topic, users' interactions are biased toward demographic homophily and ideological heterophily.

Overall, these results paint a picture where ideological echo chambers are not evident in news discussions, while we find evidence of demographic segregation.
Strikingly, this demographic segregation happens even though the demographic traits at play are not immediately available to the discussants.
Indeed, as the discussions happens online and in writing, the age, gender, and affluence of the user is a piece of latent information.
Nevertheless, these traits affect the people with whom one chooses to interact.

The most likely explanation for this result supports the theory that one's worldview is affected by the demographic group they belong to.
These worldviews, in turn, are evident in the opinions expressed through writing on online social media, which in turn drives the homophilic interaction process.
This result runs against the common narrative of the Web and the Internet as a `global village'~\citep{mcluhan1994understanding}.
Note that \citet{harteveld2021ticking} showed that higher levels of demographic segregation are associated with higher levels of affective polarization.
As such, our results support the hypothesis that the increase in affective polarization observed over the last decades might be more associated with an increasingly divided society than with online echo chambers.

Our findings also confirm that there is no ideological selective exposure in news discussions on Reddit.
While it is known that more niche subreddits might form echo chambers, with users moving within ideologically uniform communities~\cite{rollo2022communities}, our results strengthen previous analyses about the lack of ideological echo chambers in generalist Reddit communities~\citep{defrancisci2021noecho}.
Conversely, echo chambers have been reportedly observed on social media such as Twitter and Facebook~\citep{conover2011political,garimella2018political,bakshy2015exposure,an2014partisan}.
\citet{cinelli2021echo} have advanced the hypothesis that this difference between Reddit and other social media derives from their different structure: interest-based forum versus interpersonal social networks.
Future work should investigate if, on these social networks, demographic segregation also contributes to affective polarization.

Our analysis also revealed some interesting patterns about the preferences of different socio-demographic groups toward certain news topics.
It is already known that newspapers and media producers target news content to their audience, given their advertising-based financial model~\cite{elejalde2019understanding}.
Some of these preferences have also been reported in the literature.
For instance, \citet{meeter2010sports} find that gender influences news retention about business news, both age and educational attainment affect political news, and educational attainment also had an effect on science news.
It is reasonable to expect that also news consumption would be affected, and indeed, the business topic shows a marked distinction on the gender axis (with males more interested in the topic), and age has an effect on the interest in political news (older users are more interested).

In a similar spirit, \citet{lee2014newsworthy} investigate the relationship between demographic variables and the noteworthiness of news.
They find that local news is perceived as more noteworthy by women, older, and less affluent people.
In contrast, political and international news is favored by older and more educated people.
Business and sports are quite different, with the former perceived as more noteworthy by older, more educated, and more affluent people, while the latter is more noteworthy for men and white people.
Finally, entertainment is one category where younger adults have more interest than older people.
On all the demographic variables that we also measure, our results are in line with what was found by \citet{lee2014newsworthy}, who used a completely different methodology.
It is well known that different news sources have different attractiveness for different socio-demographic groups~\cite{pewresearchSectionListening}.
For instance, male and wealthy audiences prefer business news, such as the ones published on the Economist and the Wall St. Journal~\cite{pewresearchSectionDemographics}.
We see this effect in our results.

Conversely, entertainment news is often associated with a young and female audience~\cite{hamilton2011all}.
We see more mixed evidence for this pattern, with, for instance, entertainment news being more engaging for younger users, but without any effect for gender, and sports news with a similar effect for age but being more engaging for a male audience.
Healthy living and weird news, arguably two of the main soft news categories, show the opposite pattern, with a more female audience but no difference in age.

These effects suggest that, through demographic segmentation of the audience, online news might be contributing to widening the rift between demographic groups.
For instance, demographic segmentation has been observed even in political advertising on social media, where different demographic groups witness different messages~\cite{capozzi2021clandestino}.
However, this work shows that the separation between demographic groups goes beyond their interests in different news topics.
Social media users are self-sorting~\cite{harteveld2021ticking} by demographic groups, whereby interactions across groups are less likely.
The widening rift between demographic groups, whose worldviews and material conditions are growing apart, is a phenomenon that should be taken more into consideration, even in social media analysis.

This finding suggests a few research directions to overcome the limitations of the present work.
On a methodological level, this study is limited to a few demographic traits, estimated via a black-box approach.
Firstly, it would be important to develop a white-box, probabilistic assessment of demographic groups online:
having a Bayesian estimate 
would help to quantify the statistical significance of segregation patterns. %
Secondly, the inclusion of more demographic attributes beyond gender, age, and affluence, could highlight other important phenomena in online social interactions.
Furthermore, we restrict our analysis to \R{news}: other venues for online discussions might present different characteristics.
Beyond improving the methodology and broadening the scope,  %
in future works one should investigate the causal pathways that connect demographic segregation and affective polarization.
In this work, we have quantitatively compared two competing hypotheses to explain affective polarization and found that demographic segregation is more prominent than ideological echo chambers on Reddit news discussions.
More analysis of the causal connections between these important aspects of our society is urgently needed.

\clearpage
\bibliographystyle{ACM-Reference-Format}
\bibliography{references}

\end{document}